# Miniaturized wireless sensor networks

A.Lecointre*, D.Dragomirescu*, D.Dubuc*, K.Grenier*, P.Pons*, A.Aubert*, A.Muller** ,
P.Berthou, T.Gayraud and R.Plana*

*LAAS-CNRS and University of Toulouse, France
**E-mail: plana@laas.fr**
**IMT Bucarest, Romania
***E-mail**: amuller@imt.ro

*Abstract*

*This paper addresses an overview of the wireless sensor networks. It is shown that MEMS/NEMS technologies and SIP concept are well suited for advanced architectures. It is also shown analog architectures have to be compatible with digital signal techniques to develop smart network of microsystem.*

*Keywords: Miniaturized wireless sensor networks, Reconfigurable architectures, advanced signal processing.*

## 1. INTRODUCTION

It is now understood that one of the main challenge of the information and communication society deals with the ubiquitous communication. This in turns a strong need concerning both technologies, architectures and signal processing techniques.

Concerning the technologies, it can be stated that miniaturization will continue in order to increase the compactness of the modules and in order to minimize the volume, the weight of the equipment that will be more and more embedded. The miniaturization will associated with new integration technologies through heterogeneous integration techniques. This will follow the strategic evolution called "More than Moore" that is considered as the most important industrial segment for the future. In order to tackle both miniaturization and heterogeneous integration, the technologies will have to feature both sensing, communication and signal processing capabilities.

The second important issue will concern the architectures that will be needed. They will have to be able to ensure an exhaustive coverage of the mission envisioned. This will be achieved through specific networks of sensors connected by advanced communications medium. It has to be emphasized that the architectures will have to be robust and to support versatile environments. The third important issue deals with the signal processing techniques that will have to support different networks topology, different protocol of communication, a large variety of bandwidth and bit rate and to minimize the power consumption as some of them could be embedded in environment with no energy source. This paper will try to make an overview of the technologies, the architectures and the signal processing techniques that will have to be implemented in order to develop smart miniaturized wireless sensor networks. The paper will be organized as follows. Section 2 will make an overview of the technologies that could useable. Section 3 will present the architectures that exist and the one that have to be developed and/or invented when section 4 will be devoted to the signal processing techniques that will be used in these miniaturized networks.

## 2. TECHNOLOGY OVERVIEW

The last ten years have seen the emergence of MEMS/NEMS technologies that combine both bulk and surface micromachining techniques in order to develop advanced devices, circuits and systems [1]. Today it is possible to tackle the integration of MEMS/NEMS with integrated circuits through above IC and/or IN IC MEMS as plotted in figure 1 [2].

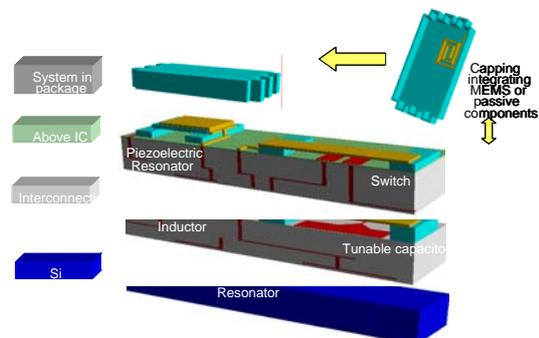

*Figure 1 : MEMS IC concept after [2]*

One of the main issue with this concept deals with the compatibility of the different technologies. In order to develop technologies that are compatible between each other it is very important to monitor, to control and to model the mechanical strain associated with the different technological processes [3]. Table 1 summarizes the different parameters that are needed to be known and the corresponding measurement techniques.

| Measurement technique | Test structure | Material | Properties | Loading | Deformation |
|---|---|---|---|---|---|
| Wafer curvature | Full wafer | All* | $\sigma$ | ***** | Profiler |
| Cantilever curvature | Cantilever | Au | $\Delta\sigma, \sigma^{**}$ | ***** | Optical profiler |
| Nano indentation | Full wafer | SiNx, Au, Cu | $E, \sigma$ | Nano indentor | Nano indentor |
| Ponctual loading | Bridge | Au | $E, \sigma$ | Nano indentor | Nano indentor |
| Ponctual loading | Membrane | SiNx, Au | $E, \sigma, \sigma_r$ | Nano indentor | Nano indentor |
| Bulge test | Membrane | SiNx, Au | $E, \sigma, \sigma_r$ | Pressure | Optical profiler |
| Vibrometry | Bridge | Au | $E, \sigma$ | Piezoelectric | Optical profiler |
| Vibrometry | Membrane | Au | $E, \sigma$ | Piezoelectric | Optical profiler |
| Electrostatic actuation | Cantilever | Au | $E(T), \sigma$ | Voltage | Optical profiler |
| Electrostatic actuation | Bridge | Au | $E(T), \sigma$ | Voltage | Optical profiler |

*Table 1 : material properties of MEMS and characterization techniques*

Using the MEMS technologies, it is now possible to fabricate a large variety of sensors (i.e pressure, thermal, mechanical, chemical and biological sensors). It has to be pointed out that the technologies have to feature some communication capabilities like antenna, filter, impedance synthesizer and phase shifter. All these functionalities are regrouped under the name of "RF MEMS". Figure 2 shows an example of two types of miniaturized MEMS based antenna that could be integrated with sensors. The antenna on the left side deals with a slot antenna that is fabricated using appropriate in IC techniques surrounding a Silicon Germanium HBT integrated process. The antenna plotted on the right side in figure 2 represents a yagi-uda suspended antenna that has the advantage of featuring an in plane radiation pattern that makes the overall architecture simpler.

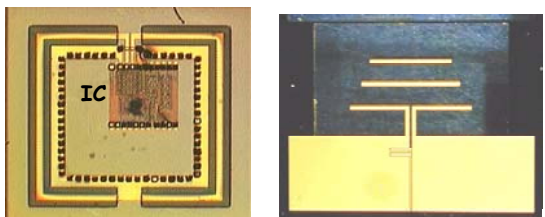

*Figure 2 : Miniaturized MEMS based antenna*

The other important issue concerning the communication deals with the fact that technological process has to support different standards. Figure 3 presents an example of a reconfigurable low noise amplifier for WLAN applications.

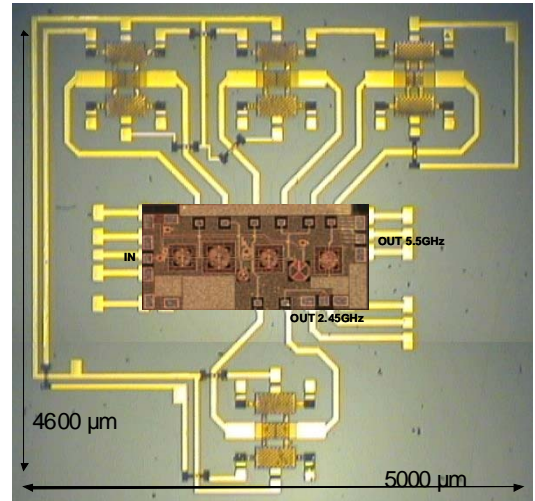

*Figure 3 : Reconfigurable LNA for WLAN applications*

One challenge that is emerging today deals with the heterogeneous integration of MEMS and IC through the system in package concept as plotted in figure 4. The main feature is to develop technologies that will afford to support smart microsystem architecture.

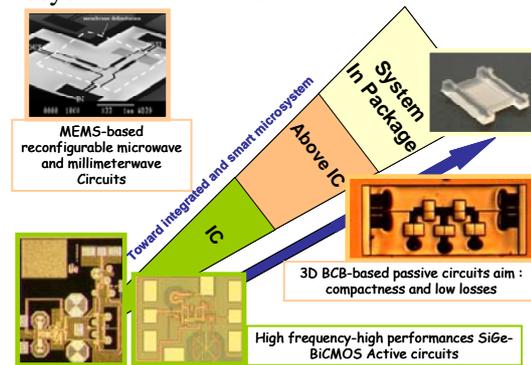

*Figure 4 : SiP concept for smart microsystem*

The next section will present the architecture that will be implemented in wireless sensors networks.

## 3. WIRELESS ARCHITECTURE

This section will aim with the architectures that will be needed to support the concept of miniaturized sensor networks. The first point deals with the wireless medium. There is frequency allocation ranging from 1 GHz to 100 GHz following the type of application targeted. In the RF range (1 GHz-6 GHz) it has to be emphasized that due to the spectrum overcrowding there is strong need concerning the interferer minimization that makes the design architecture very complicated. In order to overcome this problem, one way consists to use millimeter wave frequency range where the linearity issues are less important. It has to be pointed out that going to millimeterwave is also very interesting for some applications necessitating high bit rate. We also have to highlight that millimeterwave range turns out to very confined radiation pattern that allows the saving of electromagnetic energy. In this context, it is important to ensure a full coverage of the communication medium.

For all the architectures, it is necessary to calculate the electromagnetic budget in order to define the network organization. The following equations are presenting how to design the network.

The first point deals with the loss model that is described using the following equation:

$Lfs = 20\log_{10}(4\pi D/\lambda)$ (1)

Where D is the distance between the sensor and λ is the wave length.

The next equation deals with the noise issues that are described as the following:

$SNR = (Eb/No)*(BR/BW), N = KTBW$ (2)

Where BR is the bit rate, BW the bandwidth

Assuming equation (2), we can calculate the sensitivity of the receiver through the following equation:

$P_{rx} = N + SNR + Nf$ (3)

Where Nf is the noise figure of the receiver

Assuming equations (1), (2) and (3), we can calculate the emission power than will be needed to design the network:

$P_{tx} = P_{rx} - G_{tx} - G_{rx} + Lfs + m\arg in$ (4)

Where Gtx and Grx represent the antenna gain and margin represents a value to secure the communication (a typical value of 10 dB is usually chosen).

The next section will be devoted to the signal processing techniques that are used for wireless sensor networks.

## 4. SIGNAL PROCESSING TECHNIQUES

The first issue, we would like to present deals with a comparison of existing protocols in term of bit rate, range and power consumption [4]. Figure 5 presents a comparison of the data rate that could be achieved with different protocols.

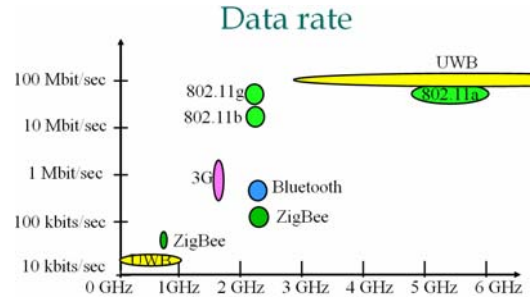

*Figure 5 : Data rate comparison for different protocols of communication*

It is shown that today a large diversity of data rate can be obtained. It has to be also pointed out that Ultra Wide Band (UWB) protocol is able to feature both low and high bit rate capabilities. It has to be noticed that for millimeterwave range UWB is attractive as it is easier to have a large useable bandwidth.

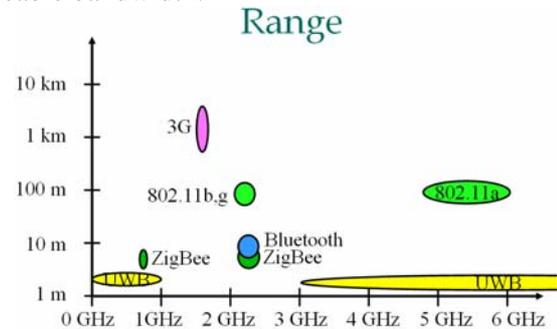

*Figure 6 : Range of communication versus the existing protocols of communication*

Figure 6 shows the range of communication calculated through equations (1,2,3,4) that shows that most of the protocols are devoted for short range communication. It has to be noticed that using millimeterwave range will be very interesting for short range communication and high bit rate. For miniaturized sensor networks, the architectures will feature numerous sensors separated by very short distance in order to have a very efficient meshing of the targeted application.

Figure 7 presents the power consumption featured by the different standard of communication. It is shown that the most efficient protocol in term of power consumption is ZigBee one.

It has to be pointed out that ZigBee protocol is not determinist and that could be a major drawback for many applications.

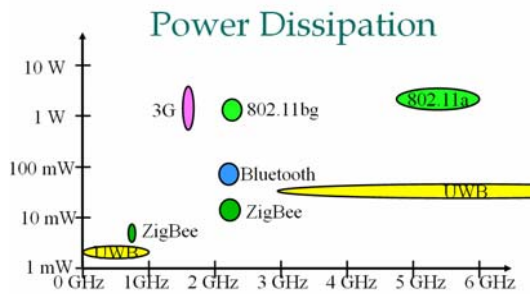

*Figure 7 : Power consumption versus the different protocols of communication*

In order to confirm this aspect, we have conducted system simulation using QUALNET software. Figure 8 presents the errors simulation for the different existing protocol of communication where we show that TDMA transmission scheme is the one featuring no error that is normal as it is a determinist scheme.

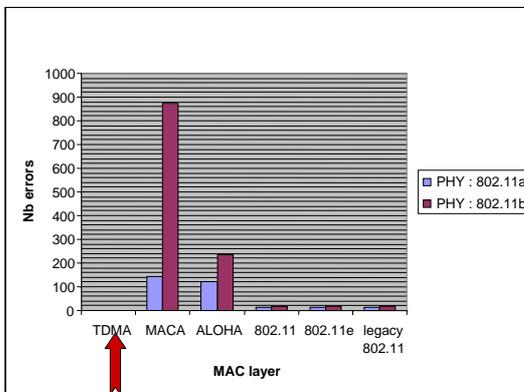

*Figure 8 : Errors simulation for different protocol of communication*

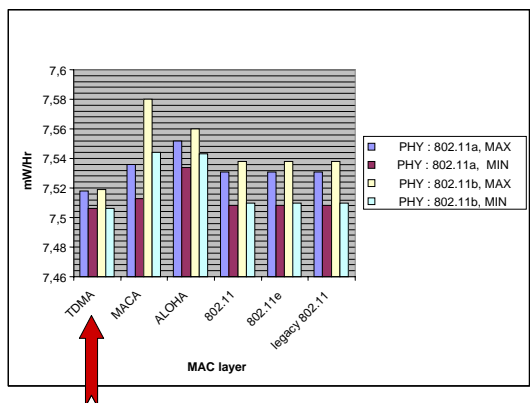

*Figure 9 : power consumption for different protocol of communication*

In figure 9, we are presenting the simulation of the power consumption exhibited by the different protocol of communication where we find that TDMA scheme presents a power consumption lower than 10mW/hour that is very important when sensor networks are abandoned and needs some autonomy. It has to be pointed out that TDMA scheme has a drawback related to the delay in the communication that has to be included in the system simulation.

## 5. CONCLUSIONS

This paper presents an overview of the technologies, the architectures and the signal processing techniques that are competing for miniaturized sensor networks. It is shown that MEMS technologies and System In Package integration are well suited technologies. It is also shown that the architectures have to be robust reliable and reconfigurable to support different standard of communication. It is also presented that TDMA is a good trade-off between data rate, range and power consumption.